\documentclass{JINST}

\input editing.sty

\newcommand{\OMISSIS}[1]{\begin{center}{$\mathbf{[\dots \: \mathrm{OMISSIS} \: \dots]}$}\end{center}}
\def\TQL{{\tt TQL}}
\def\LONE{{\tt L1}}

\def\VCOM{V_{\mathrm{PType5}}}
\def\VAVR{V_{\mathrm{PType1}}}

\def\duty{\delta_{\mathrm{cycle}}}

\def\Tup{T_{\mathrm{up}}}
\def\Tdown{T_{\mathrm{down}}}
\def\Vlow{V_{\mathrm{low}}}
\def\Vup{V_{\mathrm{up}}}

\def\Vhigh{V_{\mathrm{high}}}

\def\PPCorr{\Delta T_{\mathrm{pck}-\mathrm{peak}}}
\def\PPCorrCalc{\Delta T^{\mathrm{Calc}}_{\mathrm{pck}-\mathrm{peak}}}
\def\CIpp{\mathcal{C}T_{\mathrm{pck}-\mathrm{peak}}}
\def\CI{\mathcal{C}^I_{1,5}}

\def\OBT{t^{\mathrm{ob}}}
\def\OBTpeak{t^{\mathrm{ob}}_{\mathrm{peak}}}
\def\OBTpacket{t^{\mathrm{ob}}_{\mathrm{packet}}}

\def\folding{\mathrm{folding}}

\newcommand{\tausamplingpi}{{\tau'}_{\mathrm{sampling},i}}     % f sampilng

\def\taupck{\tau_{\mathrm{pck}}}     % tau packet
\def\Nsamplespck{\mathcal{N}_{\mathrm{pck}}}     % tau packet

\def\REBAnaver{\mathrm{N_{aver}}}
\def\REBAq{\mathrm{q}}
\def\REBAo{\mathrm{Offset}}
\def\REBAgo{\mathrm{GMF1}}
\def\REBAgt{\mathrm{GMF2}}

 \def\sky{sky}
 \def\load{reference--load}

 \def\onboard{on--board}

 \newcommand{\tausampling}{\tau_{\mathrm{sampling}}}     % f sampilng
 \newcommand{\freqsampling}{f_{\mathrm{sampling}}}     % f sampilng
 \newcommand{\Naver}{N_{\mathrm{aver}}}     % Naver
 \newcommand{\Qerr}{\epsilon_{q}}

 \newcommand{\Pone}{P_{1}} % Tsky
 \newcommand{\Ptwo}{P_{2}} % Tload

 \newcommand{\round}{\mathrm{round}}

  \newcommand{\sinc}{{\mathrm{sinc}}}

\usepackage{subfig}

\def\FIGPTYPES{
\begin{figure*}
 \centering
 \includegraphics[width=0.75\textwidth]{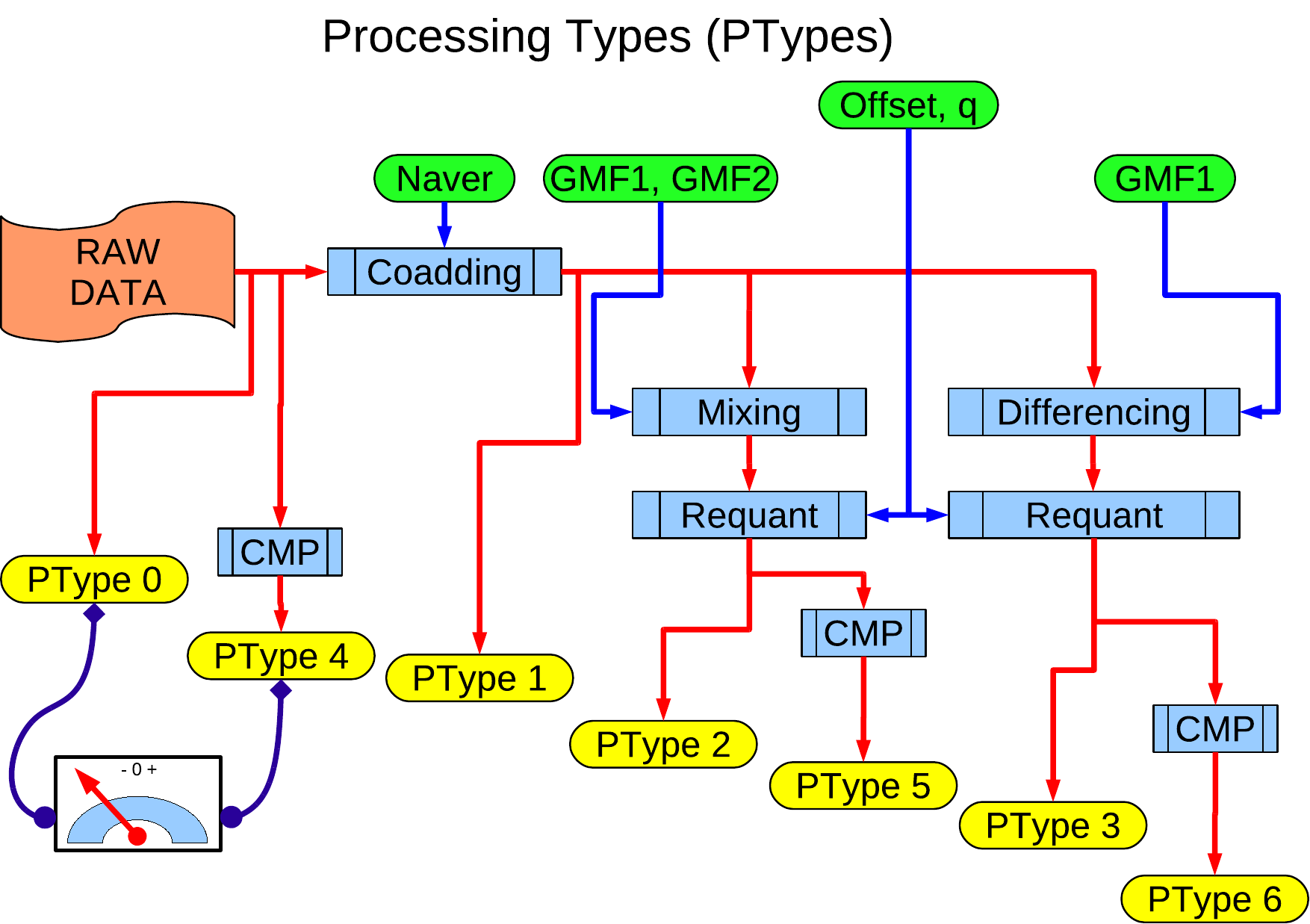}
\caption{
Schematic representation of the scientific \onboard\ processing, processing
parameters and processing types for the LFI. The diagram shows the sequence of
operations leading to each processing type: coadding, mixing, requantization
(Requant) and compression (CMP). A detailed description of the operations is
given in \cite{Fra09}.
\label{fig:ptypes}
}
\end{figure*}
} % \FIGPTYPES

\def\FIGVALIDATIONSCHEME{
\begin{figure*}
 \centering
 \includegraphics[width=0.9\textwidth]{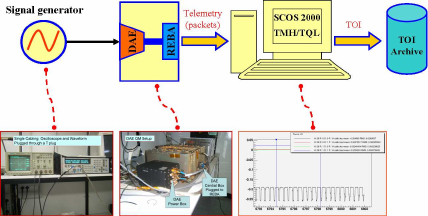}
\caption{
Hardware setup during the end-to-end tests.
\label{fig:validation:scheme}
}
\end{figure*}
} % \FIGVALIDATIONSCHEME

\def\FIGREGISTRATION{
\begin{figure*}
 \centering
 \includegraphics[width=0.8\textwidth,height=0.5\textheight]{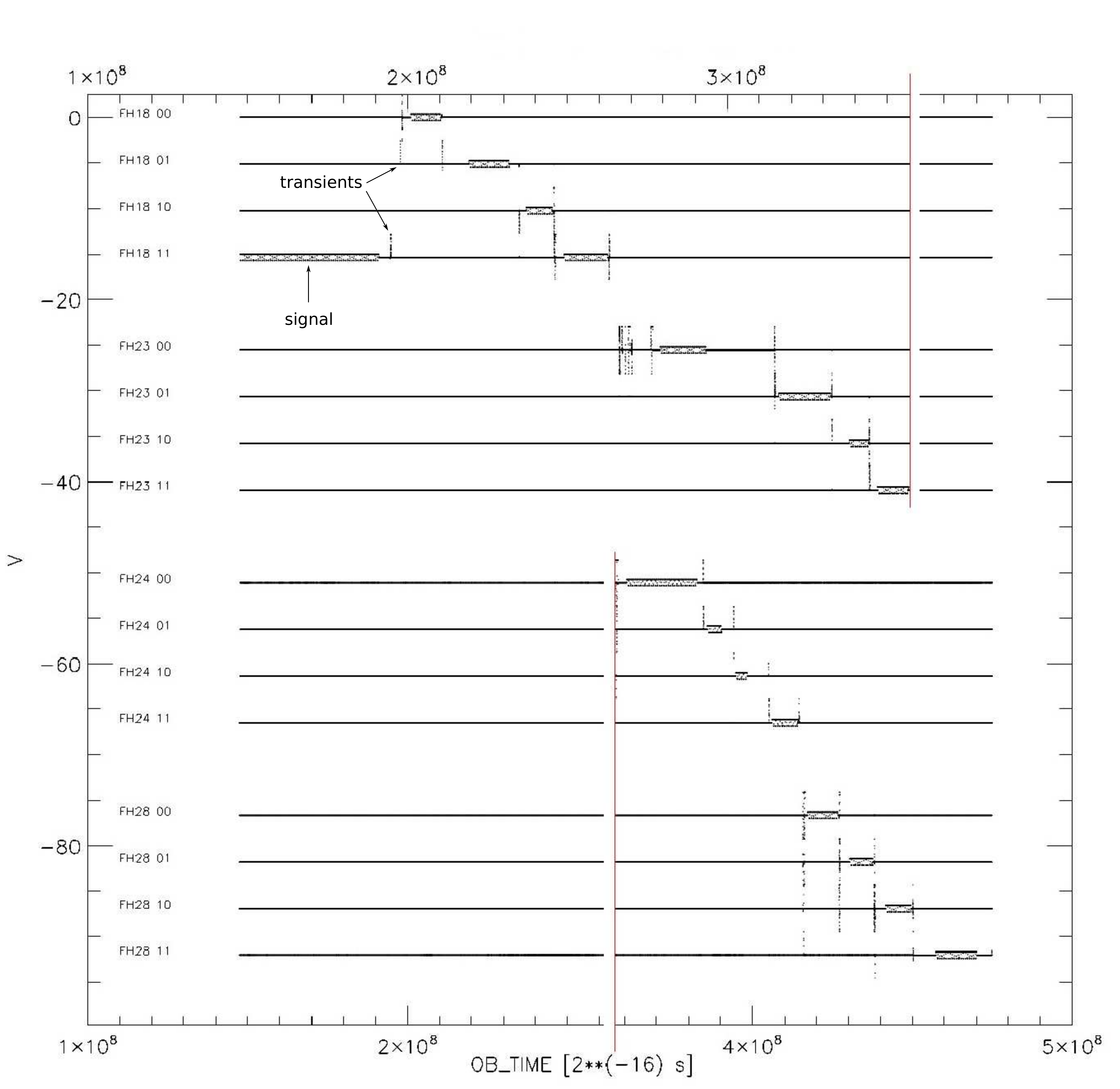}
\caption{
Simplified scheme of registration for 16 of the 44 LFI channels.
Note that the vertical scale is in volts, but each plot is vertically
shifted for graphical purposes. For the same reason the time line is split and shifted back at the level of the 
vertical red--line.
\label{fig:registration}
}
\end{figure*}
} % \FIGREGISTRATION

\def\FIGREGRESSIONTEST{
\begin{figure*}
 \centering
\subfloat[]
{
 \includegraphics[width=0.32\textwidth,height=0.32\textwidth]{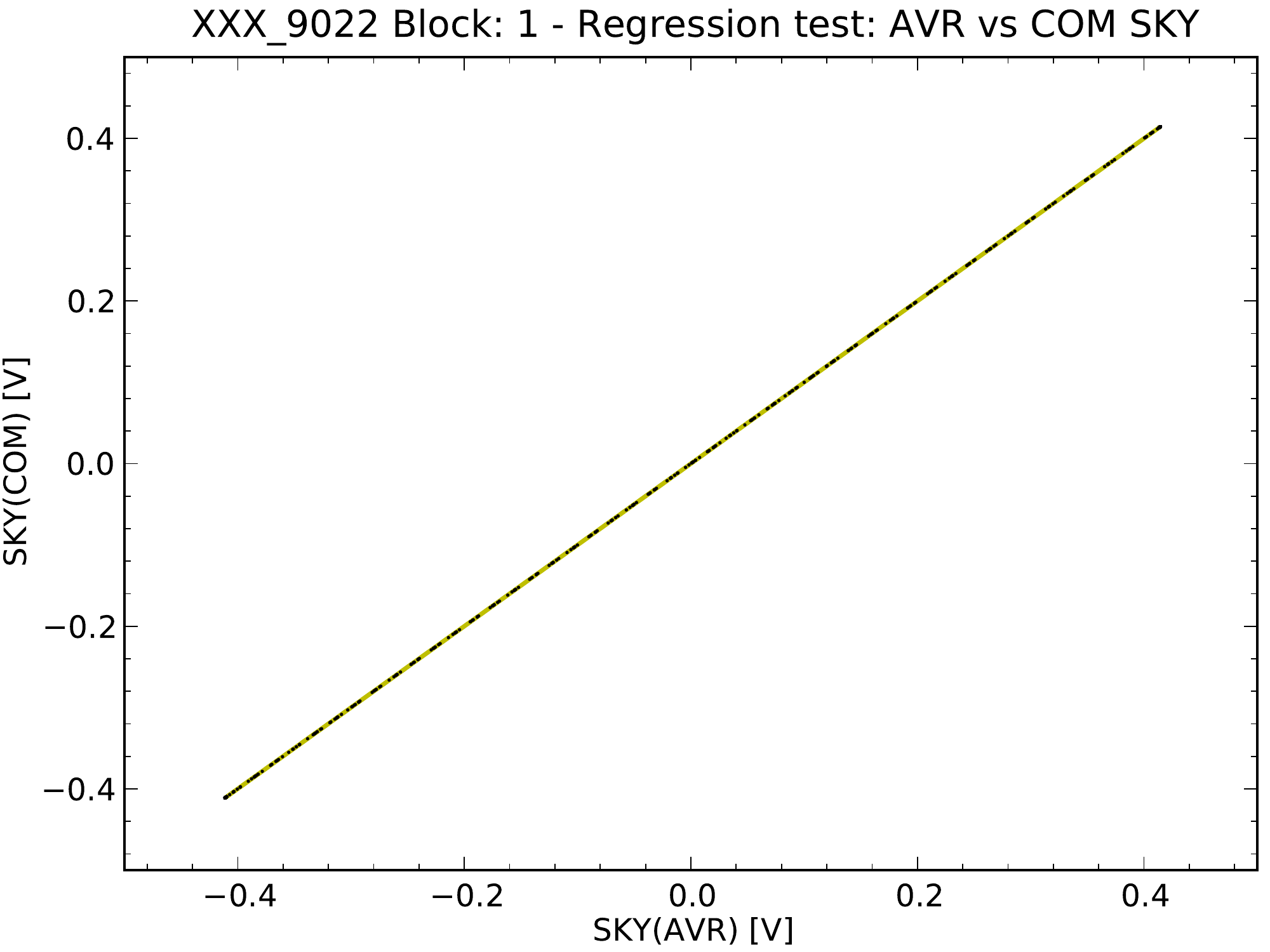}
}
\subfloat[]
{
\includegraphics[width=0.32\textwidth,height=0.32\textwidth]{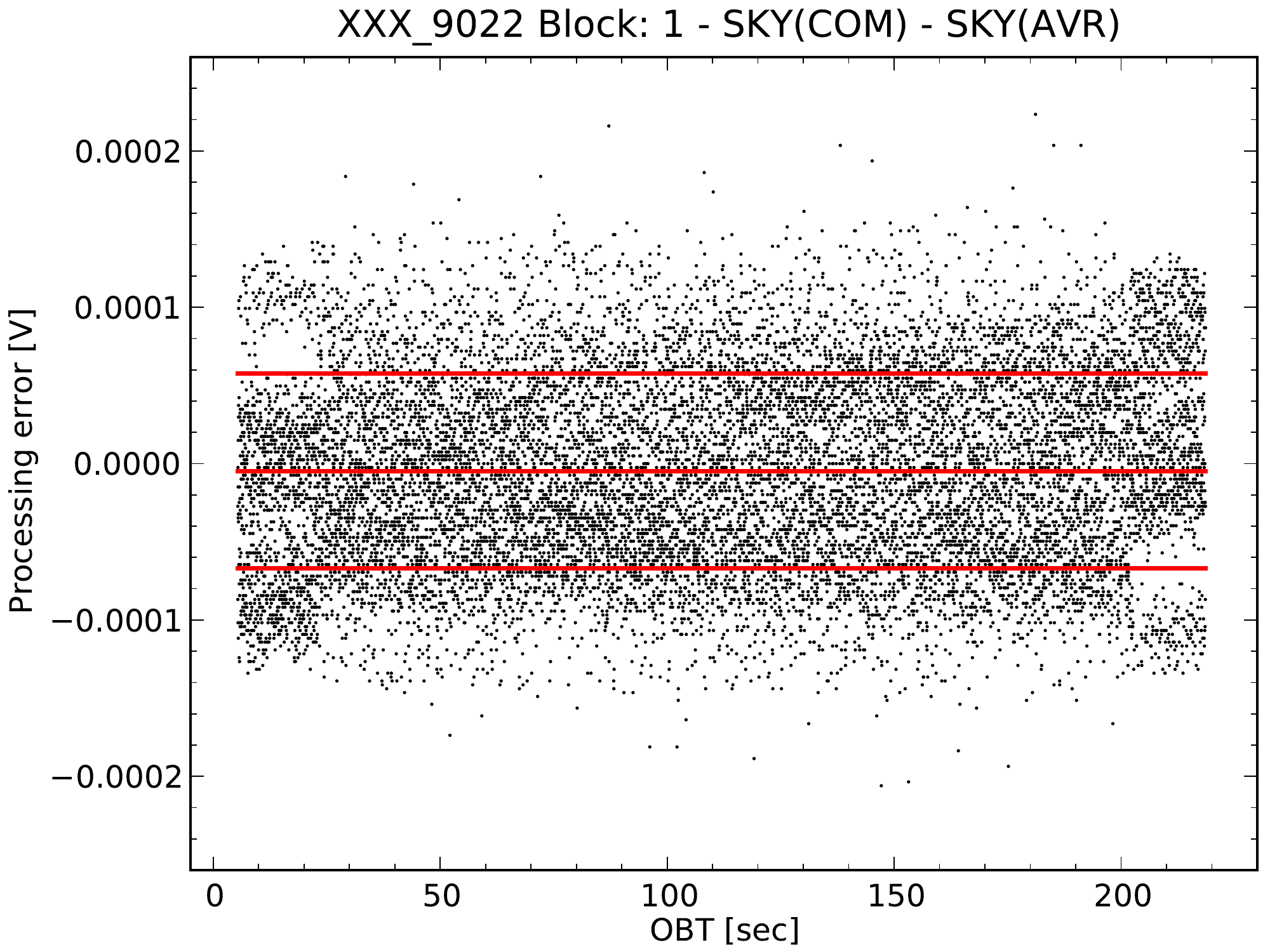}
}
\subfloat[]
{
 \includegraphics[width=0.32\textwidth,height=0.32\textwidth]{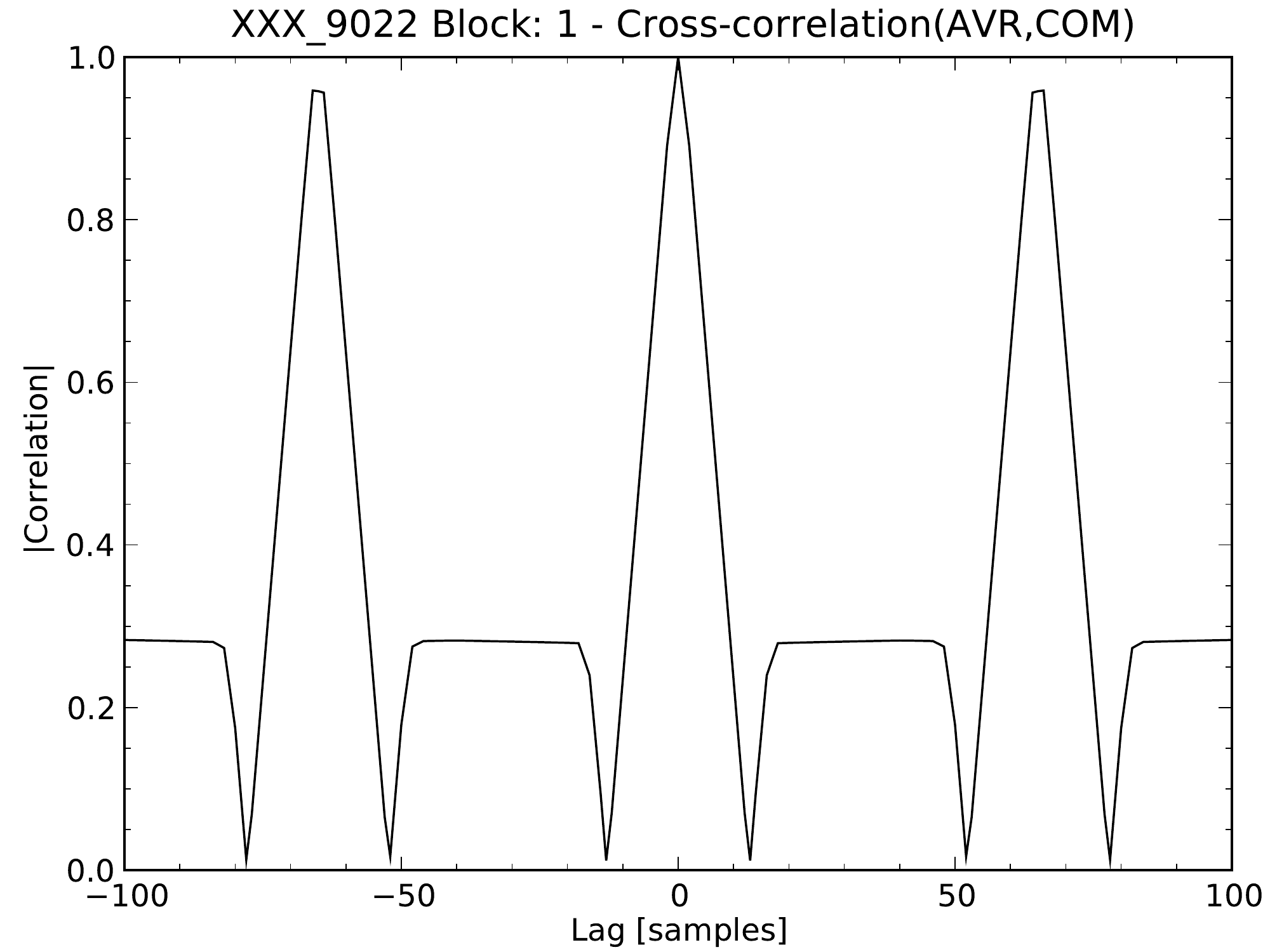}
}
\caption{
Frame a) Comparison of two different processing types (PType5 and PType1)
applied to the same data stream. Points are the original data, the yellow line is the ideal relation.
Frame b) Processing error introduced by PType5. The red lines represent the average expected processing error and the $\pm 1$ band. Note that the distribution of this error is visibly not Gaussian. Besides, digitization of data is very tiny and the distribution of noise in the region where the signal generator is off is not uniform.
Frame c) Cross-correlation test between PType 5 and PType 1.
\label{fig:regression:test}
}
\end{figure*}
} % \def\FIGREGRESSIONTEST

\def\FIGQERRORDUED{
\begin{figure}[htp!]
 \centering
\subfloat[]
{
 \includegraphics[angle=0,width=0.40\textwidth,height=0.40\textwidth]{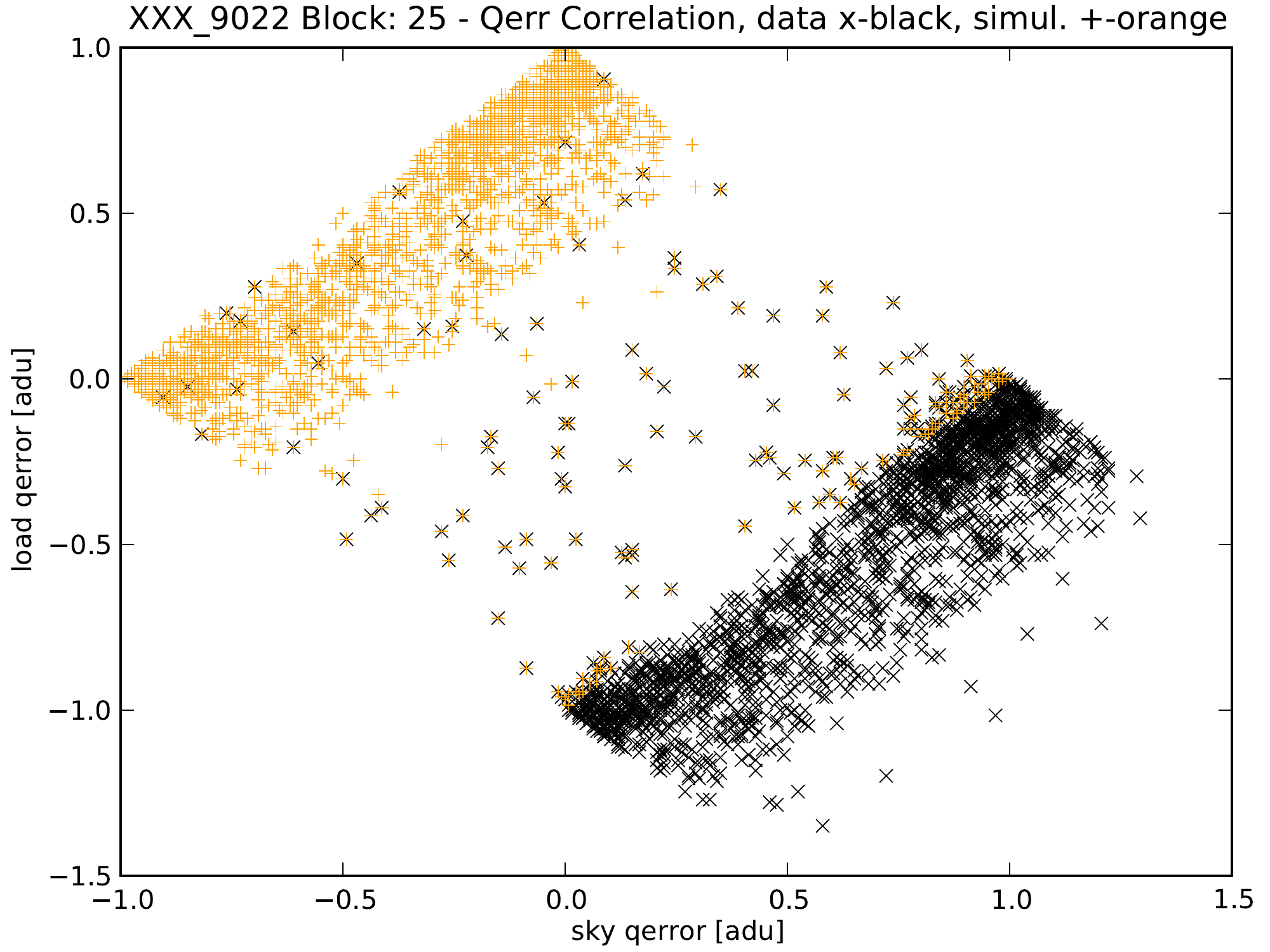}
}
\subfloat[]
{
 \includegraphics[angle=0,width=0.40\textwidth,height=0.40\textwidth]{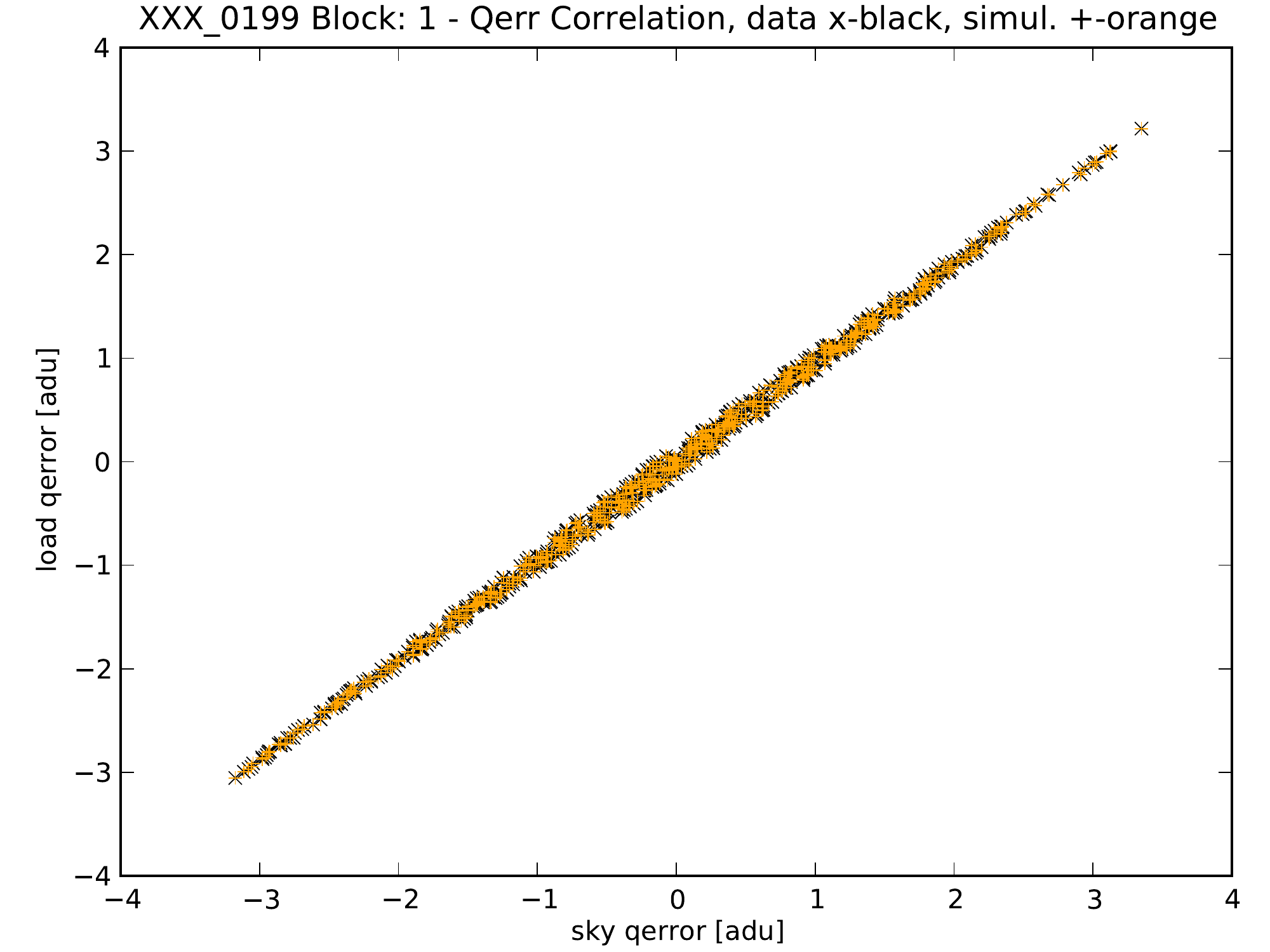}
}
\caption{
Frame a) Distribution of quantization errors for sky vs quantization errors for
load for true PType 5 data (black crosses) and the simulated processing (orange crosses).
Frame b) The same test performed on another data set after the bug on the
on-board software was corrected.
\label{fig:qerror:2d}
}
\end{figure}
} % \FIGPTYPES

\def\FIGHVSSYSTEM{
\begin{figure}[htb!]
 \centering
 \includegraphics[width=0.7\textwidth]{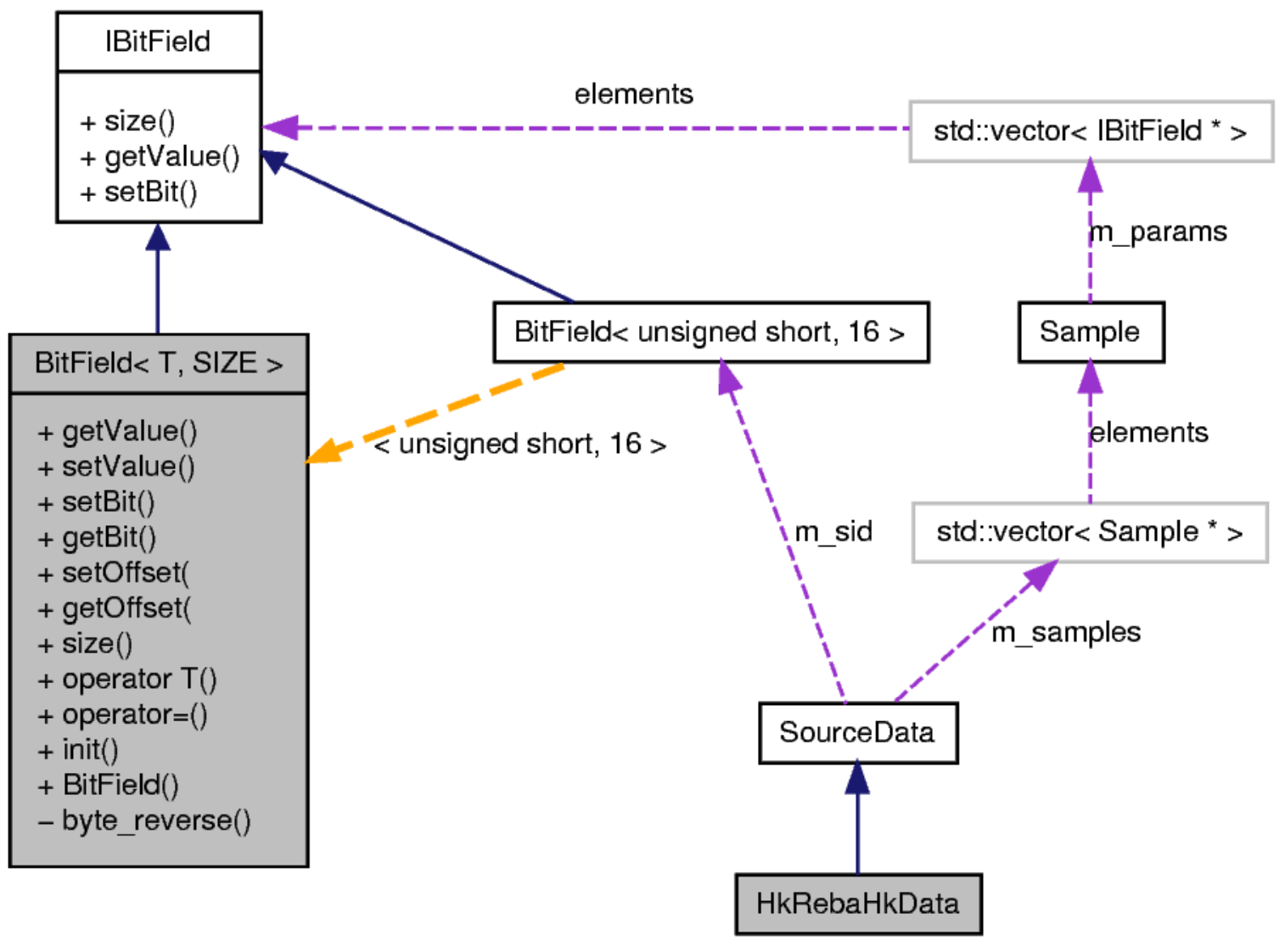}
\caption{
Class diagram of the core components of the HVS system.
\label{fig:hvs}
}
\end{figure}
} % \FIGHVSSYSTEM

\def\FIGHKERR{
\begin{figure}
 \centering
 \includegraphics[width=0.7\textwidth]{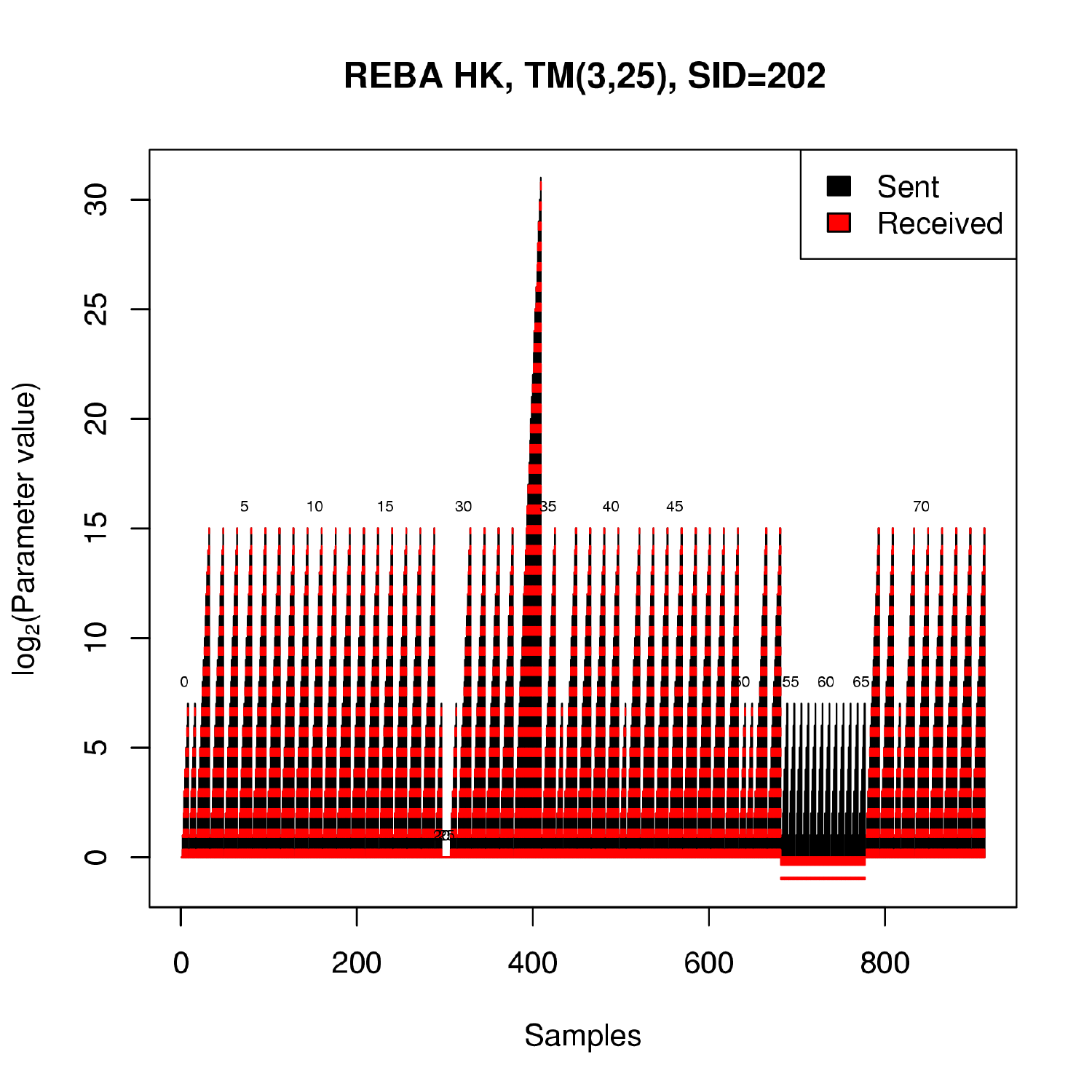}
\caption{
Comparison between the HK parameters values in the HVS generated dataset (black) with values in the
corresponding TOIs (red) for the REBA HK telemetry packet.
\label{fig:HkErr}
}
\end{figure}
} % \FIGHVSSYSTEM

\def\TABTAUSAMPLINGDIAGNOSTIC{
\begin{table*}
 \centering
 \caption{
 Diagnostic of $\tausampling$.}
\begin{tabular}{cccl}
 \hline \hline
  \multicolumn{3}{c}{{\bf Case}} & \multicolumn{1}{c}{{\bf Diagnostic}} \\
 \hline
 $\tausamplingpi$ & $\le$ & $ 0$  & repeated packets or reset in the \onboard\ clock \\ 
 $\tausamplingpi$ & $ <$ & $   \tausampling$  & overlapping of packets or erroneous decrease in $\Naver$\\ 
 $\tausamplingpi$ & $ =$ & $   \tausampling$  & the nominal condition\\ 
 $\tausamplingpi$ & $ >$ & $   \tausampling$  & gap in the acquisition or erroneous increase in $\Naver$\\ 
 \hline \hline
\end{tabular}
\label{tab:tausampling:diagnostic}
\end{table*}
} % \def\TABTAUSAMPLINGDIAGNOSTIC

\def\instOATS{$^a$}
\def\instISDC{$^b$}
\def\instIASFBO{$^c$}
\def\instSISSA{$^d$}
\def\instUNIMI{$^e$}
\def\instUNITS{$^f$}
\def\instJBO{$^g$}
\def\instCNR{$^h$}

\title{A systematic approach to the {\sc Planck} LFI end-to-end test and its
  application to the DPC Level 1 pipeline \footnote { Submitted to JINST: 24 June 2009, Accepted: 23 November
2009, Published: 29 December 2009.
 Reference : 2009 JINST 4 T12021
 DOI: 10.1088/1748-0221/4/12/T12021
  }}

\author{
M.~Frailis{\instOATS}\thanks{Corresponding Author,
e--mail:frailis@oats.inaf.it},
M.~Maris{\instOATS},
A.~Zacchei{\instOATS},
N.~Morisset{\instISDC},
R.~Rohlfs{\instISDC},
M.~Meharga{\instISDC},
P.~Binko{\instISDC},
M.~T\"urler{\instISDC},
S.~Galeotta{\instOATS},
F.~Gasparo{\instOATS},
E.~Franceschi{\instIASFBO},
R.~C.~Butler{\instIASFBO},
O.~D'Arcangelo{\instCNR},
S.~Fogliani{\instOATS},
A.~Gregorio{\instUNITS},
S.R.~Lowe{\instJBO},
G.~Maggio{\instOATS},
M.~Malaspina{\instIASFBO},
N.~Mandolesi{\instIASFBO},
P.~Manzato{\instOATS},
F.~Pasian{\instOATS},
F.~Perrotta{\instSISSA},
M.~Sandri{\instIASFBO},
L.~Terenzi{\instIASFBO},
M.~Tomasi{\instUNIMI},
A.~Zonca{\instUNIMI} \\
\llap{\instOATS}Osservatorio Astronomico di Trieste, INAF, \\
Via Tiepolo 11, 34143 Trieste, Italy \\
\llap{\instISDC}ISDC, University of Geneva, \\
ch. d'Ecogia 16, 1290 Versoix, Switzerland\\
\llap{\instIASFBO}IASF - Sezione di Bologna, INAF, \\
Via Gobetti, 101, 40129 Bologna, Italy\\
\llap{\instSISSA}SISSA - ISAS, \\
via Beirut 2-4, 34151 Trieste, Italy \\
\llap{\instUNIMI}Dipartimento di Fisica, Universit\`a degli Studi di Milano,\\
Via Celoria 16, 20133 Milano, Italy \\
\llap{\instUNITS}Dipartimento di Fisica, Universit\`a degli Studi di Trieste, \\
Via Valerio 2, 34127 Trieste, Italy\\
\llap{\instJBO}Jodrell Bank Centre for Astrophysics, The University of Manchester, \\
Manchester, M60 1QD, U.K.\\
\llap{\instCNR}Istituto di Fisica del Plasma, CNR, \\
Via Cozzi 53, Milano, Italy
\vspace{8cm}}

\abstract{The Level 1 of the Planck LFI Data Processing Centre (DPC) is devoted
  to the handling of the scientific and housekeeping telemetry. It is a critical
  component of the Planck ground segment which has to strictly commit to the
  project schedule to be ready for the launch and flight operations. In order to
  guarantee the quality necessary to achieve the objectives of the Planck
  mission, the design and development of the Level 1 software has followed the
  ESA Software Engineering Standards. A fundamental step in the software life
  cycle is the Verification and Validation of the software. The purpose of this
  work is to show an example of procedures, test development and analysis
  successfully applied to a key software project of an ESA mission. We present
  the end-to-end validation tests performed on the Level 1 of the LFI-DPC, by
  detailing the methods used and the results obtained. Different approaches have
  been used to test the scientific and housekeeping data processing. Scientific
  data processing has been tested by injecting signals with known properties
  directly into the acquisition electronics, in order to generate a test dataset
  of real telemetry data and reproduce as much as possible nominal
  conditions. For the HK telemetry processing, validation software have been
  developed to inject known parameter values into a set of real housekeeping
  packets and perform a comparison with the corresponding timelines generated by
  the Level 1. With the proposed validation and verification procedure, where
  the on-board and ground processing are viewed as a single pipeline, we
  demonstrated that the scientific and housekeeping processing of the Planck/LFI
  raw data is correct and meets the project requirements.}


\begin{document}

\section{Introduction}\label{sec:introduction}
The design and development of the LFI-DPC software follows the guidelines
provided by the ESA Software Engineering Standards \cite{ESA91}. The purpose of
these standards is to guarantee as much as possible the quality of the software
produced, in terms of compliance with the user and software requirements,
stability, accuracy, efficiency and commitment to the project schedule.

A fundamental stage in the software life cycle is the software verification and
validation. According to the ESA standard \cite{ESA95}, ``verification'' means
the act of reviewing, inspecting, testing, checking, auditing, or otherwise
establishing and documenting whether items, processes, services or documents
conform to specified requirements. ``Validation'' is the process of evaluating a
system or component during or at the end of the development process to determine
whether it satisfies specified requirements. Hence, validation consists of an
``end-to-end'' verification, where known inputs are provided to the system and,
after correct processing, the outputs are analyzed and cross-checked with the
expected results.

In this paper we present the approach adopted by the LFI DPC for the end-to-end
tests of the LFI Level 1 software by describing methods, test procedures and
results of the validation. The Level 1 (L1) of the LFI processing is formed by
three subsystems:
\begin{itemize}
\item A Real-time Assessment system (RTA): based on SCOS
  2000\footnote{www.egos.esa.int/portal/egos-web/products/MCS/SCOS2000/}, the
  ESA generic mission control system and Electrical Ground Segment Equipment
  (EGSE), it is a system for the real-time monitoring of the housekeeping
  telemetry, in order to verify the overall health of the instrument and detect
  possible anomalies. It receives telemetry packets directly from the instrument
  and provides TCP/IP and CORBA interfaces to communicate with external devices
  and software.
\item A TeleMetry Handler system (TMH): developed by the ISDC (Integral Science
  Data Center) team, on the basis of the user requirements defined by the LFI
  DPC, the TMH system is the core of the L1 pipeline. It receives in real-time
  scientific (SCI) and housekeeping (HK) telemetry packets from the RTA system,
  then it decodes their content, reconstructs the time of the SCI and HK samples
  and creates time series (also called Time Ordered Information, TOIs) for each
  data source and acquisition mode.
\item A Telemetry Quick-Look system (TQL): mainly developed to perform an
  interactive quick analysis of the LFI data, it provides a set of graphical
  tools to display the LFI scientific data and the satellite HK data and
  calculates quick statistics and fast Fourier transforms.
\end{itemize}
A more detailed description of the LFI-DPC Level 1 software can be found in
\cite{Fra09}.

The development and release of the Level 1 software has followed the model
philosophy adopted for the development stages of the LFI instrument (and in
general for the space projects \cite{ECSS08}). Each model of the software
implemented the additional functionalities necessary to support the tests,
calibration and operations of the corresponding model of the instrument. First,
a Bread-Board Model of the Level 1 was released in 2002, followed by a
Demonstration Model \cite{Pas05}. For the Qualification Model of the
instrument, providing a prototype fully representative of the instrument
functionality \cite{Men06}, a Qualification Model version of the Level 1 was released in 2005. A
Flight Model release has followed to support the calibration and
integration tests of Planck/LFI. The last Level 1 software model, the Operations
Model, integrating all the Planck Science Ground Segment interfaces,
has been finally released to support the Planck nominal flight operations.

The tests have been focused on the validation of the TMH/TQL system and the
customizations applied to the standard distribution of SCOS 2000. The tailoring
of SCOS concerns the task handling the reception of the telemetry from an
external source, the task sending the telemetry to the TMH and the definition of
the Mission Information Base. A first run of the tests has been performed
on the Qualification  Model of the TMH/TQL. Subsequently, tests have been
repeated on the Flight Model, after solving the bugs identified in the
previous release.

The main testing scheme used is a black-box testing, where known inputs have
been provided to the pipeline and the results of the processing have been
compared with the expected outputs. To test the scientific telemetry processing,
a digital signal generator has been connected directly to the analog gates of
the LFI acquisition chain in order to sample a signal with known
properties. Metrics have been applied to the output of the TMH/TQL to assess the
proper reconstruction of the input signal. For the housekeeping telemetry
processing, software have been developed to insert known parameter values into
an archive of real LFI HK packets \cite{Fra06}.

\section{ The LFI software  verification and validation}

Verification \& Validation (V\&V), within the Planck/LFI DPC context, has to be
intended as Scientific Software Verification and Validation. This means that the
main subject of V\&V is the quality of the scientific product which may be
obtained with the software delivered and installed at the DPC. The
validation process assures that all scientific software meet their
functional and performance requirements according to a given set of
validity criteria.

First, a Software Integration, Verification and Validation Plan (SIVVP) has been
developed by the LFI DPC to define the basic principles, methods and procedures
that shall be followed for the software verification and validation at the
acceptance point \cite{Pop06}, in order to assure
\begin{itemize}
\item the validation of each software block taking into account its functional
  and performance requirements;
\item the verification of each software component at specific points of its life
  cycle;
\item the accuracy of the software against the accuracy stated by the software
  procedures;
\item that the software satisfies all the requirements defined in the
  corresponding User Requirement Document (URD).
\end{itemize}

Acceptance testing should require no knowledge of the internal workings of the
software. Hence, the general procedure for the verification and validation of
the software, at the acceptance point, is a {\bf black-box} test. In such type
of tests, the software is treated as a ``black-box'' whose internals cannot be
seen \cite{Som06}. The validation process consists of the comparison of the results of
each software component with the results expected by the user for a given
predefined input dataset through the application of predefined validity
criteria. When an incremental delivery approach is used, acceptance tests only
address the user requirements of the new release, since regression tests are
performed in system testing. For each test, validity criteria are given as a
list of Test Pass/Fail Criteria (TP/FCs).

\section{The LFI acquisition chain and the \LONE\ pipeline}\label{sec:LFI:acqusition}

\FIGPTYPES

A detailed description of the LFI \onboard\ acquisition chain and the LFI \LONE\
processing is given by \cite{Fra09}. In this work, we use a simplified model of the \LONE\
pipeline in order to present the test methods and procedures applied for the
end-to-end tests of the \LONE\ software.

In this model, data in form of time series are generated from various SCI
and HK {\em sources} (i.e. analog detectors, sensors, registers)
\onboard\ the spacecraft and read by the Digital Acquisition Electronics (DAE)
which samples the analog signals at discrete times. The analog output of the
instrument read by each scientific ADC is either a sequence of interleaved
\sky\ and \load\ measures, or a sequence of measures of only \sky\ or only
\load, depending on the status of the phase--switches associated with each
radiometer of the LFI instrument.

The DAE is polled at regular intervals by the on-board processor, the REBA,
which runs a pre-processing pipeline, reducing the amount of
scientific data generated by the instrument and storing the data in telemetry
packets to be sent to ground. A set of 7 extraction points are defined along the
REBA pipeline, in order to recover valuable diagnostic data. Each point
corresponds to a {\em Processing Type} (PType) applied by the REBA on the
scientific data (see figure~\ref{fig:ptypes}). The REBA software can apply
concurrently two different PTypes to a single LFI channel. So, in nominal
conditions, the instrument will generate just PType 5 data from all of its 44
detectors and short chunks of PType 1 data from a single detector in turn. 

In particular, in PType 5 three main processing steps are applied
\cite{Fra09}. First, a double difference is performed between two averaged \sky\
and \load\ samples, $\overline{S}_{sky}$ and $\overline{S}_{load}$:
\begin{eqnarray}
   \Pone & = & \overline{S}_{sky} - \REBAgo \cdot \overline{S}_{load} \\
   \Ptwo & = & \overline{S}_{sky} - \REBAgt \cdot \overline{S}_{load}
\end{eqnarray}
where $\REBAgo$ and $\REBAgt$ are two different gain modulation factors. Then,
the two values obtained are requantized converting them into two 16-bit signed integers:
\begin{equation}
  Q_i = \round\left[ \REBAq \cdot \left( P_i + \REBAo \right) \right]
\end{equation}
Finally, the REBA performs a loss-less adaptive arithmetic compression of the
data obtained in the previous step. Hence, the REBA pipeline is tunable through
a set of {\em processing parameters} or {\em REBA parameters}: $\REBAnaver$ (the
number of consecutive sky or load samples that are averaged), $\REBAq$,
$\REBAo$, $\REBAgo$ and $\REBAgt$ \cite{Fra09}.

The \LONE\ ground software receives the telemetry packets generated by the
instrument. In particular, the scientific telemetry is processed by the TMH/TQL
system, which has to handle properly each PType in order to regenerate
the time series acquired on-board. The exact steps of the \LONE\ pipeline processing
depend on the data source and the PType of each telemetry packet.

\section{End-to-end test objectives and methods}\label{sec:tests}
To validate the Demonstration Model of the \LONE\ software, developed before the
integration and delivery of the LFI instrument, software were developed in order
to generate simulated HK telemetry \cite{Zac03} and scientific telemetry
\cite{Mar01, Fog03}.  The scope of end-to-end (E2E) tests, instead, is to assess
the proper coverage of the requirements by the \LONE\ software and validate its
operations by using, as much as possible, the data generated by the testing
campaign of the instrument, in particular data gathered from the Qualification
Model and the Flight Model of the instrument.  The reason is the need to test
the software in the most realistic conditions, by performing real instrument
operations that allow testing of various operative aspects of the pipeline
design. The usage of simulated data has been limited only to cases in which it
was not possible to perform a complete E2E test.

Tests have been ordered according to a functional classification of the
requirements, consisting of:
\begin{enumerate}
\item Handling of raw TM packets:
  \begin{enumerate}
  \item communication with SCOS,
  \item storing of raw packets,
  \item proper interpretation of primary and secondary headers,
  \item proper commutation of packets according to their purpose,
    source, service, size and timestamp,
  \item hexadecimal dump of packet content,
  \item other packet related services;
  \end{enumerate}
\item Handling of time information;
\item Decompression, decoding and reconstruction of scientific packets content;
\item Decoding and reconstruction of HK packets content;
\item TOIs generation;
\item Graphical displays;
\item On-the-fly software analysis;
\item Other services.
\end{enumerate}

There are overlaps between some of the classes. In particular, the first 4
classes are strictly connected, allowing a minimization of the number of
tests. Classes from 6 to 8 can be tested by comparing their output directly with
the TOIs rather than the input signals. Test procedures have been defined to
cover one or more test cases of the \LONE\ E2E test plan and formalized into a
set of test controls sheets.

\subsection{Scientific data processing: testing method and setup}

Figure~\ref{fig:ptypes} suggests a possible testing strategy, i.e. {\em
  comparison testing}.  This is based on the comparison of data taken with
different PTypes. In the figure, a small gauge is connected to PType 0 and
PType 4 data. The gauge represents the \LONE\ software receiving PType 0 and
PType 4 data from the same data source (detector), processing and comparing
them. Given that the compression/decompression is a completely loss--less
process, a correct result would be PType 4 data, after decompression, to be
identical to the corresponding PType 0 data.

More generally, even an {\em absolute testing} is required to assure that data
are properly acquired, processed, received, reconstructed, displayed and stored
by the pipeline as a whole. This is obtained by acquiring a signal of well known
properties (period, shape, amplitude, voltage levels, duty cycles) at the DAE
analog gates, processing it through the whole system and comparing it with the
output by using a suitable set of metrics. Depending on the type of test,
identity between input and output or statistical agreement among them have been
used as a criteria to assess the success or failure of the test. As a by
product, this kind of testing is able to detect potential problems in the
\onboard\ acquisition electronics and software.

\FIGVALIDATIONSCHEME

Figure ~\ref{fig:validation:scheme} represents the complete acquisition chain of
LFI during the E2E tests. The signal is injected through an oscilloscope
probe at the input of a DAE pin corresponding to the detector to be tested. The
signal is generated through a stabilized, digital function generator, and it is
monitored by a digital oscilloscope connected to the signal generator by a high
impedance probe. Monitoring is required to asses that the signal produced by
the function generator matches the required characteristics. In some tests, the
same signal has been distributed to different inputs by a signal distributor.

To analyze the test data, it would have sufficed to correlate the output signal
with the input signal, but practically it was difficult to record the input signal
separately. So, the statistical properties of the output have been compared to
the properties of the input measured by the
oscilloscope and characterized by the following parameters: 
the shape, the period $T$, the duration of the
high state $\Tup$, the duration of the low state $\Tdown$, the duty cycle $\duty
= \Tup/T$, the low state voltage $\Vlow$, the high state voltage $\Vup$, and the
peak-to-peak amplitude $A = \Vup - \Vlow$.
 %
% \FIGINPUTSIGNAL
 %
The source of the signal shall ensure that all of these parameters are known
accurately, a good S/N (at least 20), that the signal fits the range of voltages
acceptable to the input gate at which it is connected, that it is stable, within
the quantization step of the ADC converter, and that $\Vlow$, $\Vup$, $\Tup$,
and $\Tdown$ are adjustable within a wide range of values.

For each test, the general procedure has been the following: 
\renewcommand{\theenumi}{\roman{enumi}}
\begin{enumerate}
\item the signal generator has been plugged to the DAE input corresponding to
  the channel to be tested;
\item the signal generator has been set-up for the type of signal needed for
  the specific test. The oscilloscope has been used to check the signal
  properties;
\item telecommands have been sent to the DAE to set the proper acquisition
  setup;
\item the DAE acquisition has been started keeping the generator off,
\item after about 5~seconds the generator has been switched on for the time
  required by the test;
\item the generator has been switched off and after about 5~seconds the acquisition
  has been stopped.
\end{enumerate}

In this procedure, the steps in which the signal generator is switched on and
off have been used to allow a semiautomatic identification of the start and end
time of each test within the timelines generated by the \LONE\ software. The
setup of each test has been recorded by documenting the cabling, the signal
generator setup, keeping screenshots of the SCOS displays and archiving the log
generated through the TQL where each test phase has been annotated. The
subsequent data analysis has been carried out off--line.

\section{Data analysis procedures}

\subsection{Software}
An analysis toolkit, {\tt OCA} (On-board Computing Analysis), implemented in the
Interactive Data Language (IDL) and in C++, with additional off-the-shelf
freeware libraries and applications, has been developed to analyze the test
data. OCA provides simple analysis methods and automated reporting. The use
of this toolkit assures repeatability of test analysis.

In particular, OCA allows the processing of the raw telemetry archive generated
by the TMH together with the intermediate products of the pipeline and the
corresponding log files. It provides functions to: decode 
the packets content, in particular the On-Board Time (OBT) and the acquisition parameters stored
in the tertiary header of the scientific packets; scan the TOI archive and
split tests into frames; detect pulses of known shapes within packets or TOIs;
identify the source packet from which a given sample was extracted; perform
comparisons between the TOIs and corresponding scientific packets. Moreover, it
simulates the conversion of PType 1 data into PType 2 data, to verify the
correct processing of these data types \cite{Mar09}.

\subsection{On-line and off-line analysis}

Depending on the scope of the test, data analysis was performed on-line and
off-line. On-line analysis has been carried out for all of the live operations
of the \LONE\ software.  For instance, we tested on-line the \TQL\ function
which generates an hexadecimal dump of the packets by comparing it with the same
function performed by SCOS, the function which monitors
the REBA parameters currently in use, the graphical displays and the FFT
computation. Most of the other functionalities, such as OBT and TOI
reconstruction, have been tested off-line.

The first group of tests have covered the basic functionalities. In particular,
we checked that all the TM packets are received and stored, that each packet
header is properly handled in the packets archive, the proper interpretation of
primary, secondary and tertiary headers, that packets are properly registered,
i.e. grouped according to the data source and processing type and converted into
TOIs. In this group of tests, the checking of the TOI creation was focused on
testing the proper mapping between the TM packets and the raw samples contained
in the TOIs.

\FIGREGISTRATION

The test checking for proper registration of scientific packets is illustrated
in figure~\ref{fig:registration}.  In that test, the acquisition was kept running
while the signal generator was connected in turn to each analog input
of the DAE. The data analysis looks for erroneous activation of unplugged channels
and/or lack of activation of the plugged ones.  It is evident that the presence
of a signal in one of the TOIs excludes signals in others (apart from short
transients, corresponding to the plugging/unplugging operations) and that there
are no holes in the data.

\subsection{Signal reconstruction}\label{sec:ptype:reconstruction}

A more advanced test procedure and analysis was required to test the OBT
reconstruction and the conversion to physical units. Ideally, those two elements
of the processing should have been tested separately. In practice, time
handling is connected to the identification of features in the signal, such as
the period between subsequent peaks or the reconstruction of the phase
information, which are obviously connected to a proper signal reconstruction.

After having identified the region where the signal generator is ``on'', a
Lomb-Scargle Periodogram (LSP) is applied to assess the period of the signal as
measured by the OBT time scale.  The LSP has the advantage, over the FFT, of
being robust against perturbations introduced by erroneous inclusions of short
no-signal zones or interruptions in the data.  The LSP is automatically scanned
looking for the peak with the highest power, giving a first approximation of the
period. The period is further refined by fitting the LSP peak with a $\sinc(x)$
with two free parameters: the difference in the peak period $\delta p$ with
respect to the approximate period and the peak normalization. The fitting
procedure estimates also the error in the two free parameters. The error in
$\delta p$ is taken as the error in the period estimate.

The phases and amplitude determination depends on the wave shape. 
For square waves, $\Vlow$, $\Vhigh$, amplitudes and phases are determined by using a $\chi^2$
fitting combined with a phase--folded diagram. 
The $\chi^2$ is computed between the model calculated for a given tentative
phase, $\Phi$, and the signal. A fine-grained grid of phases is explored and the
minimum $\chi^2$ is used to assess the best phase. 

The model for the triangular wave assumes that the transition between the
growing and the decreasing ramp is instantaneous. The peak-to-peak amplitude for
the triangular wave is determined by the moments of the cumulative distribution
function of the samples.  In particular

  \begin{equation}
   A = \sigma_{\mathrm{triangular}} \sqrt{12}.
  \end{equation}

 \noindent
In this way, the determination of the peak-to-peak
amplitude is decoupled from the determination of period and phase. 
The phase is again determined by the $\chi^2$ method used for the square waves.

\subsection{PType comparison}
Since the REBA is able to process data from the same detector with two different
processing types in parallel, this functionality was used to perform a
comparison between the TOIs of the same signal obtained from two PTypes. An
example is given in figure~\ref{fig:regression:test}, where data acquired with
PType 1 have been compared with the same data acquired with PType 5. The example
concerns a test where a square wave with a period of 1~second, peak-to-peak amplitude
of about 0.84 V and duty cycle $\duty = 0.25$ has been used; phase switching was
left off, so each resulting TOI contains only sky samples.  The figure
covers the analysis performed on data from feed-horn 28, radiometer 0 and
detector 0 only. The processing parameters used are: $\REBAnaver$=126, $\REBAgo$=1, $\REBAgt$=-1,
$\REBAo$=0, $\REBAq$=1.

\FIGREGRESSIONTEST

In particular, the first frame in figure~\ref{fig:regression:test} shows a
correlation plot of PType 5 (COM) versus PType 1 (AVR) data. As it is possible
to see, the correlation is perfectly linear. A more refined test is the
evaluation of the processing error (or quantization error) introduced by the
PType 5 transformations. It is defined as:

\begin{equation}
\Qerr = \VCOM - \VAVR
\end{equation}

\noindent
where $\VAVR$ and $\VCOM$ are, respectively, the measures obtained acquiring the
data at the same time by using both the 1 and 5 processing modes. The error,
compared to the expectation from processing parameters, is plotted in the second
frame of figure~\ref{fig:regression:test}. It shows the effect of the digitization
which, in this case, is very tiny. Some structure in the noise is clearly present,
especially at the left and right side where the signal was off. Moreover, the
noise does not follow a Gaussian distribution.

An additional test performed is the cross-correlation index, $\CI =
\rho_{1,5}(0)$, between PType 1 and PType 5. We evaluated the cross-correlation
for a range of time lags. The cross-correlation is calculated by shifting the
TOI obtained from PType 5 data with respect to the one obtained from PType 1
data by a lag of $\Delta t$, taking $\Delta t$ = 0 as the case for overlapping
PType 5 with PType 1. 

The results obtained always showed $\rho_{1,5}(\Delta t)
\leq \rho_{1,5}(0)$. The third frame of figure~\ref{fig:regression:test} shows
an example of such case for one of the tests, where we have obtained $1
- \CI = 1.81 \cdot 10^{-8}$.

\subsection{OBT reconstruction}
Testing for proper handling of time information is critical since time series
are matched by comparing samples with the same OBT.  In particular the matching
of the OBTs is used to correlate scientific time series with HK time series.
OBTs are also used to correlate scientific data acquired by using different
PTypes from the same detector and estimate the quantization error $\Qerr$.

In particular, when testing the OBT reconstruction, we consider the following
type of anomalies: gaps between the last sample of a packet and the first sample
of the subsequent one; overlaps between the last samples of a packet and the
first samples of the subsequent one; the presence of duplicated packets;
possible errors in estimating the time scale (sampling step) and the zero point
of the OBT.

The test procedures for the OBT reconstruction consist in:
\begin{enumerate}
\item checking whether the sampling rate derived from OBT is consistent with $N_{aver}$;
\item checking whether there are defects (holes) in the OBT;
\item verifying that holes (if present) in the OBT are just due to acquisition stops;
\item measuring periods, duty-cycles of signals and checking if they are
  consistent with the periods imposed by the signal generator;
\item checking that the correlation between the OBT of a packet and the OBT in the
  corresponding TOIs is correct;
\item correlating two different PTypes applied to the same signal.
\end{enumerate}

\subsubsection{Anomalies in the sampling rate}

The sampling period for PType 1, 2, 3, 5  and 6 data shall be 

 \begin{equation}
\tausampling = \frac{\REBAnaver}{8192} \; \mathrm{seconds}.
\end{equation}

 \noindent
For any pair of consecutive samples $[i, i+1]$, where $i=0$
denotes the first sample in the TOI, the OBT should be  a monotonically
increasing quantity; so, denoting with $\OBT$ the OBT and having

 \begin{equation}
  \tausamplingpi =  \OBT_{i+1} - \OBT_i 
\end{equation}

 \noindent
then for any $i$, $\tausamplingpi \equiv \tausampling$.
Hence, a simple histogram of $\tausamplingpi$ allows discovering the presence of
anomalies, i.e.  cases in which $\tausamplingpi \ne \tausampling$.
Table~\ref{tab:tausampling:diagnostic} gives a summary of the possible anomalies and their most likely 
source. Gaps in the timelines are allowed when the acquisition is stopped.
\TABTAUSAMPLINGDIAGNOSTIC
The test is complemented by cross-checking the OBT in the packet secondary
header and the corresponding OBTs in the TOIs.

In an early test, a problem discovered by this method was an inconsistent
distribution of $\tausampling$ for PTypes 1 and 2 when the phase switch was
off. They were a factor of 2 too large.  This occurred because of an incorrect
handling of the OBTs by the TMH in this instrument configuration.  After fixing
the problem the test was successfully repeated.  No defects or unexplained gaps
in the OBT have been found when the phase switch was on.

Another example of a problem discovered and corrected by looking at an erroneous
OBT reconstruction was the case of an early attempt to carry out acquisition of
PType 2 and 5 together.  In that case, the TMH did not register properly the
PType 5 data since the data of both PTypes were stored in the same TOI instead
of storing them into separate timelines.  Since PType 5 packets are generated at a
rate of about 1/6 of PType 1, the OBT apparently jumped back about every 6
PType 1 packets.

\subsubsection{Packet-Peak correlation}
Packet-Peak correlation is a method to assess the use of the OBT as a way to
correlate packets to events occurring in different time lines. The measure of
Packet-Peak correlation is assessed by measuring the time interval between
the first peak contained in the packet, $\OBTpeak$, and the time stamp of the
packet, $\OBTpacket$.

For square waves, the peak position is defined as the OBT for which 
$\folding(\OBTpeak) = \duty/2$
then a square wave ``belongs'' to a packet if 
$\OBTpacket \le \OBTpeak.$

Then, the packet-to-peak correlation
$$\PPCorr = \OBTpeak - \OBTpacket $$
can be easily predicted when the period of production of packets and the
period of the square wave is known. In particular, 
for uncompressed data, packets are generated with period:
\begin{equation} 
    \taupck = \frac{\Nsamplespck \REBAnaver}{\freqsampling} \; \mathrm{seconds},
\end{equation}
where $\Nsamplespck$ is the number of samples in the packet: 490 for any uncompressed PType, except
PType 1 for which $\Nsamplespck=245$. The deviation of $\PPCorr$ from the
predicted value is measured by:
 \begin{equation} 
    \CIpp = \frac{\left|\PPCorr - \PPCorrCalc\right|}{\taupck},
 \end{equation}
\noindent
where $\PPCorrCalc$ is the $\PPCorr$ predicted. The test fails if $\CIpp > 1/(10^3\REBAnaver) \approx 10^{-6}$.
In no cases a packet--peak correlation worst than $\CIpp < 10^{-11}$ have been detected.

\subsubsection{Period determination}
An erroneous reconstruction of the OBT would result in an erroneous estimate in
the period of the signal as measured by using the LSP. Given that the OBT of the
packets is given in seconds, with a 16~bits accuracy for the fractional
part, the maximum error in determining the period should be 
of the order of $10^{-4}$~seconds.
 
To check for this case we did various tests with different combinations of
signal periods, amplitudes and $\REBAnaver$. Periods in the measured waveforms
have been fitted by using the LSP together with phase folding as explained in
Sect.~\ref{sec:ptype:reconstruction}. Typically, accuracies better than
$9\times10^{-5}$~sec have been obtained.

\subsection{ADU conversion}

The use of a signal generator allows conversion from ADU (Analog
to Digital Unit) to physical units (Volts) to be tested. A linear fit between the fitted $V_{fit}$
and the generator values $V_{generator}$,
\begin{equation}
V_{fit} = \textrm{Intercept} + \textrm{Slope}\cdot V_{generator},
\end{equation}
has been performed. Pearson statistics have been used as a way to assess
linearity. The fit has been
performed either comparing the High levels, the Low levels or both.
\noindent
The ideal case would be:
\begin{equation}
\textrm{Intercept} = 0, \textrm{Slope} = 1,\textrm{Pearson} = 1 .
\end{equation}
\noindent
Typically, we obtained $|\textrm{Intercept}| < 5\times10^{-3}$,
$|1-\textrm{Slope}| < 10^{-2}$ and $|1-\textrm{Pearson}| < 10^{-4}$.

\FIGQERRORDUED

\subsection{Quantization errors}

Comparing PType 5 and PType 1 data allowed discovering anomalies in the
quantization error. Tests have shown that the differences between PType 1 and
PType 5 are small and compatible with the expected quantization error (depending
on the REBA processing parameters applied). However, in a first run of these
tests, in some cases it was noted that the distribution of quantization errors
did not follow the expectation exactly, since they were not symmetrically
distributed around zero.

A more detailed analysis is shown in Fig~\ref{fig:qerror:2d}a, where the
processing leading to PType 5 generation from PType 1 data has been carried out
according to the documented on-board algorithm. The figure shows a scatter plot
of sky vs. load processing error. It is possible to see that the expected
processing error is not distributed as the real one. After the analysis,  it was
concluded that it was due to a bug in the on-board software, when rounding
negative values in the quantization step. The bug was subsequently corrected and
a second run of the tests has shown that the quantization error now follows the
expectation (Fig~\ref{fig:qerror:2d}b). 

\subsection{Decompression}
PType 2 and PType 5 data where acquired together.  Data from PType 5 processing
are compared with data from PType 2 processing.  No differences were revealed
between PType 2 and PType 5 data.  The compression/decompression procedure
passed this test.

\section{Housekeeping processing validation using a known pattern}\label{sec:hk:validation}

Unlike the case of the scientific telemetry, it wasn't possible to inject known
signals into the HK packets directly through the DAE/REBA chain. The HK
parameters are heterogeneous and include temperature sensor values, current
consumptions, register values, switch statuses. Different parameters are
grouped by hardware unit, sampling frequency and service purpose and stored in
the same HK packet. 

The structure of the HK telemetry packets strictly follows the rules of the ESA
Packet Utilization Standard (PUS). According to the PUS, each field in a telemetry HK
packet is either a parameter field containing a parameter value or a structured
field containing several parameter fields organized according to a set of
rules. The PUS permits the structure of a HK packet to be defined by specifying,
for each parameter within the packet, the offset with respect to the packet
primary header, the parameter type code and format code and the offset of the
subsequent samples of the parameter.

The HK packet structure definitions are provided in an Interface Control
Document (ICD) of the instrument and then mapped in the SCOS Mission Information
Base (MIB), a set of ASCII tables containing all the monitoring parameters
characteristics and location within each telemetry packet and the telecommands
characteristics. Both the SCOS system and the TMH/TQL system of the LFI-DPC
import the packet structure definitions from the MIB tables.

In order to test the decoding and reconstruction of the HK packets content, an
Housekeeping Validation System (HVS) has been developed to manipulate the binary
representation of an HK packet with the goal of generating packets with known
parameter values starting from a set of real HK packets.

\subsection{The HVS system}

The HVS system has been developed to test the correct processing of each HK
packet type. Packets of the same type have an identical structure and are
identified by 4 fields: the Application ID (APID), the service Type and Subtype,
contained in the packet primary and secondary headers, and the Structure ID
(SID), contained in the first two bytes of the data segment.

For each HK packet type and parameter within it, the HVS system creates a
predefined pattern of values. It iterates over all the samples of the parameter,
setting to 1 one bit a time.  Hence, for each sample, only a single bit of a
single parameter has value 1 while all the other bits have value 0. This implies
that for a given HK packet type, each parameter in turn takes increasing power
of 2 values. One of the purposes of this pattern is to verify that in the
TMH/TQL system the offset and the length of each parameter has been correctly
defined.

\FIGHVSSYSTEM

Figure~\ref{fig:hvs} shows the class diagram of the components of the HVS system
which build an HK packet structure. The data segment (source data) of a packet
starts with the SID field, used to identify the packet type. Then a sequence of
samples follow. Each sample is composed of a sequence of different parameters. A
parameter is represented by a bit field (BitField template class) specifying the
size in bits of the parameter, its raw value type and the offset of the
parameter within the sample. The BitField class inherits from a pure abstract
class, the IBitField class, which is used to iterate over all parameters,
independently of their concrete type, to set each bit value.

The HVS system is able to increase the number of packets in the input dataset,
introducing new packets with coherent OBT, sample time and source
sequence count values and keeping constant the proportion of each packet type
within the dataset. With this feature, we have produced a dataset corresponding
to an acquisition lasting 6 days. This was necessary to apply the pattern of
values to packets with a low frequency (64 seconds) and containing a large
number of HK parameters.

The dataset generated by the HVS system is then processed by the TMH in order to
generate the HK timelines. For each packet type, the parameter values of the
input dataset are compared with the corresponding TOIs generated by the TMH in
order to highlight possible discrepancies. 

\subsection{HK tests results}

\FIGHKERR

Tests performed with the HVS system have highlighted an error in the handling of
two HK telemetry packets types, the REBA HK packet and the REBA Diagnostic HK
packet, which have a common structure. The problem is shown in figure
\ref{fig:HkErr} for the REBA HK packet: it represents a plot of the HK parameter
values in the dataset generated by the HVS system (black) and the corresponding
TOIs generated by the TMH system (red). The parameters are ordered according to
their position in the packet (denoted by the number on the top of each peak).
The ``sample'' axis is the location in the test sequence where that sample is
tested with the given value. The problem with the registration of the HK
parameters is evidenced by the anomalous distribution of red points around the
sample 700. This test has proved that there was a discordance between the ICD
describing the packets structure for LFI and their mapping in the SCOS~2000 MIB
\cite{Mar08}. After this analysis, the offsets of the involved parameters were
corrected in the MIB tables by the instrument team and successfully checked with
a rerun of this test.

\section{Conclusions}\label{sec:res:concl}
The Level 1 software system of Planck/LFI, which is a critical component of the Planck
science ground segment, has been validated by applying the ESA standards for
software verification and validation. The tests have been designed to recreate
the most realistic operational conditions, using telemetry directly generated by the
instrument, whenever possible. An additional effort was needed to develop
testing and analysis tools devoted to the end-to-end tests. Some relevant bugs have
been identified in the first run of the tests, performed on the Qualification
Model of the software, and rapidly fixed. The Flight Model of the Level 1 has
successfully passed the end-to-end validation tests and has been adopted for the
entire LFI calibration tests campaign and the subsequent Planck integration
tests.   

\vspace{4mm}

\acknowledgments Planck is a project of the European Space Agency
with instruments funded by ESA member states, and with special
contributions from Denmark and NASA (USA). The Planck-LFI project
is developed by an International Consortium lead by Italy and
involving Canada, Finland, Germany, Norway, Spain, Switzerland,
UK, USA. The Italian contribution to Planck is supported by the
Italian Space Agency (ASI). The US Planck Project is supported by
the NASA Science Mission Directorate.

\end{document}